\documentclass[conference]{IEEEtran}
\usepackage{graphicx}
\usepackage{booktabs}
\usepackage{array}
\usepackage{subcaption}
\usepackage{url}
\usepackage{listings}
\usepackage{color}
\usepackage{multirow}
\usepackage{paralist} % for inline list
\usepackage{amsmath}
\usepackage{enumitem}
\usepackage{caption}
\usepackage{float}
\usepackage{wrapfig}

\usepackage{flushend}

\begin{document}
\title{Keep Your Nice Friends Close, but Your Rich Friends Closer --
Computation Offloading Using NFC}
\author{
\IEEEauthorblockN{Kathleen Sucipto$^\ast$, Dimitris Chatzopoulos$^\ast$,  Sokol Kosta$^{\pm\dagger}$, and Pan Hui$^\ast$}
    \IEEEauthorblockA{$^\ast$HKUST-DT System and Media Lab, %Department of Computer Science and Engineering
The Hong Kong University of Science and Technology\\
$^\pm$CMI, Aalborg University Copenhagen, Denmark\\
$^\dagger$Sapienza University of Rome, Italy \\
ksucipto@connect.ust.hk, dcab@cse.ust.hk, sok@cmi.aau.dk, panhui@cse.ust.hk
\vspace{-.5cm}
}   
}

\maketitle

\begin{abstract}
The increasing complexity of smartphone applications and services necessitate high battery consumption but the growth of smartphones' battery capacity is not keeping pace with these increasing power demands. To overcome this problem, researchers gave birth to the Mobile Cloud Computing (MCC) research area.
In this paper we advance on previous ideas, by proposing and implementing the first known Near Field Communication (NFC)-based computation offloading framework.
This research is motivated by the advantages of NFC's short distance communication, with its better security, and its low battery consumption.
We design a new NFC communication protocol that overcomes the limitations of the default protocol; removing the need for constant user interaction, the one--way communication restraint, and the limit on low data size transfer.
We present experimental results of the energy consumption and the time duration of two computationally intensive representative applications: \textit{(i)} RSA key generation and encryption, and \textit{(ii)} gaming/puzzles.
We show that when the helper device is more powerful than the device offloading the computations, the execution time of the tasks is reduced. Finally, we show that devices that offload application parts considerably reduce their energy consumption due to the low--power NFC interface and the benefits of offloading.
\end{abstract}

\section{Introduction}
Mobile users nowadays are demanding more and more functionalities and sophisticated services to be supported by their devices.
Unfortunately, the more complex a functionality or a service becomes, the more energy it typically consumes.
To this end, developers are starting to feel a lot of pressure, due to the fact that advances on battery technology are not able to keep pace with the energy demands of modern applications.
Even though there has been much interest in enhancing smartphones' lithium-ion battery capacity, any significant improvement would take a long time to occur \cite{cite1}. Eventually, a new generation of rapid-charging smartphone batteries, like nanodot-based batteries~\cite{cite2}, will be developed, but these batteries may not be publicly available until late 2016 or 2017~\cite{cite3}.
To cope with these limitations, developers struggle to carefully implement the heavy tasks of an application so to not drain the battery very quickly, while still offering the desired services to the users.

Recently, with the advent of cloud computing, one popular adopted solution is \textit{computation offloading}~\cite{Balan:2002:CCF:1133373.1133390}; a method where resource-intensive computations are executed remotely in one or more powerful machines known as \textit{offloadees} or \textit{surrogates}.
Not only is this considered to be a potential way to conserve battery power, it can also reduce the total execution time \cite{MAUI,CLONECLOUD,kosta2012thinkair}.
So far, many computation offloading frameworks using Bluetooth or Wi-Fi have been developed. However, despite having reasonable bandwidth or data rate, they are still facing some limitations, such as: the interference of other WiFi or Bluetooth devices, the Internet connection requirement in the case of MCC, which implies higher energy needs and higher delay~\cite{MAUI}, and the difficulty in detecting available nearby devices for computation offloading through WiFi-direct or Bluetooth~\cite{SERENDIPITY_}. 

In this paper we advance on the previous ideas and implement for the first time, to the best of our knowledge, a mobile offloading framework over NFC, a wireless protocol based on Radio Frequency Identification (RFID) \cite{want2006introduction}.
NFC enables devices with a distance of less than 10 cm to exchange small amounts of data \cite{coskun2013professional}. 
The majority of recent Android devices already implement this functionality.
Although the bandwidth of NFC is typically about 50-340 times smaller than Bluetooth and Wi-Fi~\cite{cite5}, NFC's short range and operational characteristics provide several advantages compared to Bluetooth and Wi-Fi. For instance, NFC guarantees low interference and has lower energy consumption~\cite{Chang:2010:NNP:1805161.1805165, cite7}.
Many researchers believe that NFC has promising potential for future applications; with many research groups working to enhance and apply this technology~\cite{ozdenizci2010nfc,coskun2013professional}.
Several of NFC's advantages, especially its \textbf{low--energy} demands, its intrinsic security that comes from the short--range communication, and its current active development, make this technology highly suitable for computation offloading.

We envision a not--so--distant future reality where \textit{Internet of Things} will surround us in every aspect of our life, with objects interacting with each--other in a myriad of ways. 
We believe that smart tables or smart desks, like Microsoft Surface Tabletop~\cite{cite9} e.g., will be available in every home, every office, and every bar.
Combining the computational capabilities of these smart surfaces with the potential of NFC communication, we can easily see the benefits that this technology enables when people put their NFC--capable smartphones on the surface, as they normally do today.
Making use of the NFC offloading framework, a mobile device can transfer all the heavy computations to the smart surface, reducing its energy consumption and quite often improving the execution time of such heavy operations.

In this paper we investigate and bypass the current limitations of the NFC protocol in order to design and build a fully working offloading framework.
In more detail;
\begin{inparaenum}[(\itshape 1\upshape)]
	\item First, we modify the default NFC communication protocol to make it work without user intervention and to be able to exchange data bidirectionally;
	\item We measure several characteristics of our new NFC protocol, such as bandwidth and latency, showing that bidirectional transmission of small data is feasible;
	\item Then, following the same techniques as previous MCC frameworks---such as remote method invocation, we implement an NFC-based computation offloading framework, which is made possible thanks to our new NFC communication protocol;
	\item We identify typical smartphone applications that can benefit from NFC computation offloading; 
	\item Finally, we evaluate the performance of our framework using two representative applications, showing that not only offloading is possible but also convenient in terms of reduced energy consumption and improved execution performance on the mobile device that requests offloading.
\end{inparaenum}

The rest of the paper is organised as follows:
In Section~\ref{sec:related} we position our paper with respect to relevant works in this area;
in Section~\ref{sec:des} we present the design and implementation of our several NFC communication protocols alongside with the details of the offloading framework;
in Section~\ref{sec:eval} we compare our different NFC protocols to choose the best one;
in Section~\ref{sec:univ} we present the final NFC offloading framework built on top of the optimized NFC protocol;
in Section~\ref{sec:exp} we evaluate our framework through two representative applications and discuss the experimental results;
finally, in Section~\ref{sec:concl} we present the conclusion remarks and future work.

\section{Related Work}\label{sec:related}
Much work has been done on exploring the concept of computation offloading and applying it on mobile devices.
MAUI~\cite{MAUI} supports code offloading from smartphones to nearby servers or devices in order to minimize energy consumption.
The results show significant energy savings when Wi-Fi is used, while when using 3G the results are not very satisfactory.
CloneCloud~\cite{CLONECLOUD} aims to benefit directly from the cloud, transforming a mobile application by migrating parts of its execution to a virtual machine on the cloud.
ThinkAir~\cite{kosta2012thinkair} combines the advantages of these frameworks and works with Wi-Fi and 3G offloading to nearby or remote surrogates.
Furthermore, ThinkAir allows for the computational power to be dynamically scaled up or down on the cloud, enabling high levels of flexibility for the developers.

More recently, Serendipity~\cite{SERENDIPITY_} introduces the concept of mobile--to--mobile offloading in an environment with intermittent connectivity.
This system is capable of conserving energy and increasing computation speed of low--power devices when these offload heavy computations to more powerful ones.
Nevertheless, similar to other previous frameworks, Serendipity relies on Wi-Fi, and the paper does not specify how to search for and detect the available devices that are willing to help. OPENRP advances in this direction by collecting data from interactions between mobile users and building reputation scores per mobile user and application type \cite{7509388}. 
Honeybee \cite{Fernando2013} is an offloading framework for mobile computing on Bluetooth channels.
Without having to rely on Wi-Fi, it guarantees connectivity as long as other mobile devices equipped with Bluetooth are available as well.

Seen that mobile code offloading has already become well accepted and its advantages have been acknowledged by many authors, researchers lately have been focusing on building more solid frameworks that consider long neglected aspects, such as security, fault--tolerance, caching, etc.~\cite{Zhang:2014:CED:2685048.2685057}.
Gordon et al.~\cite{Gordon:2015:AMA:2742647.2742649} replicate mobile applications, which are split into execution phases in mobile servers, efficiently selecting the proper replica to proceed in the next phase, in order to improve the end users' quality of experience.  
Bouzefrane et al. propose a security protocol for authentication between NFC applications and proximal cloudlets motivated by the fact that NFC applications can be computationally demanding~\cite{6834974}.

However, the current state of their project is quite preliminary and does not present any real implementation. Furthermore, the design is quite limited, due to the fact that it requires the user to constantly tap on the device: once when offloading the security computation and another tap when receiving the result.
Conversely, in this work we have redesigned the NFC communication protocol to eliminate the need for user intervention, which enables a convenient and automated offloading process.
NFC-based computation offloading does not need to consider most of the problems that traditional offloading frameworks face.
For example, the low--range communication capabilities of NFC eliminate the need for data encryption, which of course is an overhead that current frameworks have to deal with \cite{7130873}.
Moreover, our architecture allows the mobile devices to automatically connect with the powerful offloadee entities, since they will be in close NFC proximity, eliminating the need for long registration process as presented in~\cite{kosta2012thinkair} or intentionally neglected as in other works.

\begin{table}[t]
   \begin{center}
    \begin{tabular}{ m{2 cm} m{1.7 cm} c c}
    \toprule
    Name & CPU & Memory & OS \\
    \midrule
    Xiaomi Mi 3 & Quad-core 2.3 GHz Krait 400 & 2 GB & Android 4.4.4 \\ 
    \midrule
    Samsung Galaxy Note 3 & Quad-core 2.3 GHz Krait 400 & 3 GB & Android 4.4 \\
    \bottomrule
	\end{tabular}
   	\caption{Devices used in our experiments. \vspace{-0.5cm}}
	\label{tab:devices}
	\end{center}
\end{table}

\section{Design and Implementation}\label{sec:des}

In this section we describe the requirements and the steps followed to implement a functional computation offloading framework between smartphones over the NFC communication channel.
First, we describe briefly the limitations of the current NFC hardware and software interfaces, which make it difficult to build a fully functional NFC offloading framework.
Then, we describe the steps we undertook to overcome such limitations and build the framework.
In the rest of the paper we refer to the device asking for computation offloading as the \textit{main device} or the \textit{offloading device}, while to the device executing the offloaded computation as the \textit{offloadee device} or simply the \textit{offloadee}.
In Table~\ref{tab:devices} we show the specs of the two devices used in our experiments; a Xiaomi Mi~3 as the main device and a Samsung Galaxy Note~3 as the offloadee device.

\subsection{Preliminaries}

\subsubsection{Limitations of NFC Android API}
The basic implementation of the NFC in Android is based on \textit{Nfc Data Exchange Format (NDEF)}\footnote{http://developer.android.com/reference/android/nfc/tech/Ndef.html}. 
The data transmission can only be unidirectional, which of course does not allow for successful computation offloading since the result of the offloaded task cannot be sent back to the offloading device.
Moreover, to trigger the transmission of an NDEF message (\textit{NdefMessage}) the Android operating system requires the user to tap on the smartphone's screen.

\begin{figure}[t]
\centering
    \includegraphics[width=0.75\columnwidth]{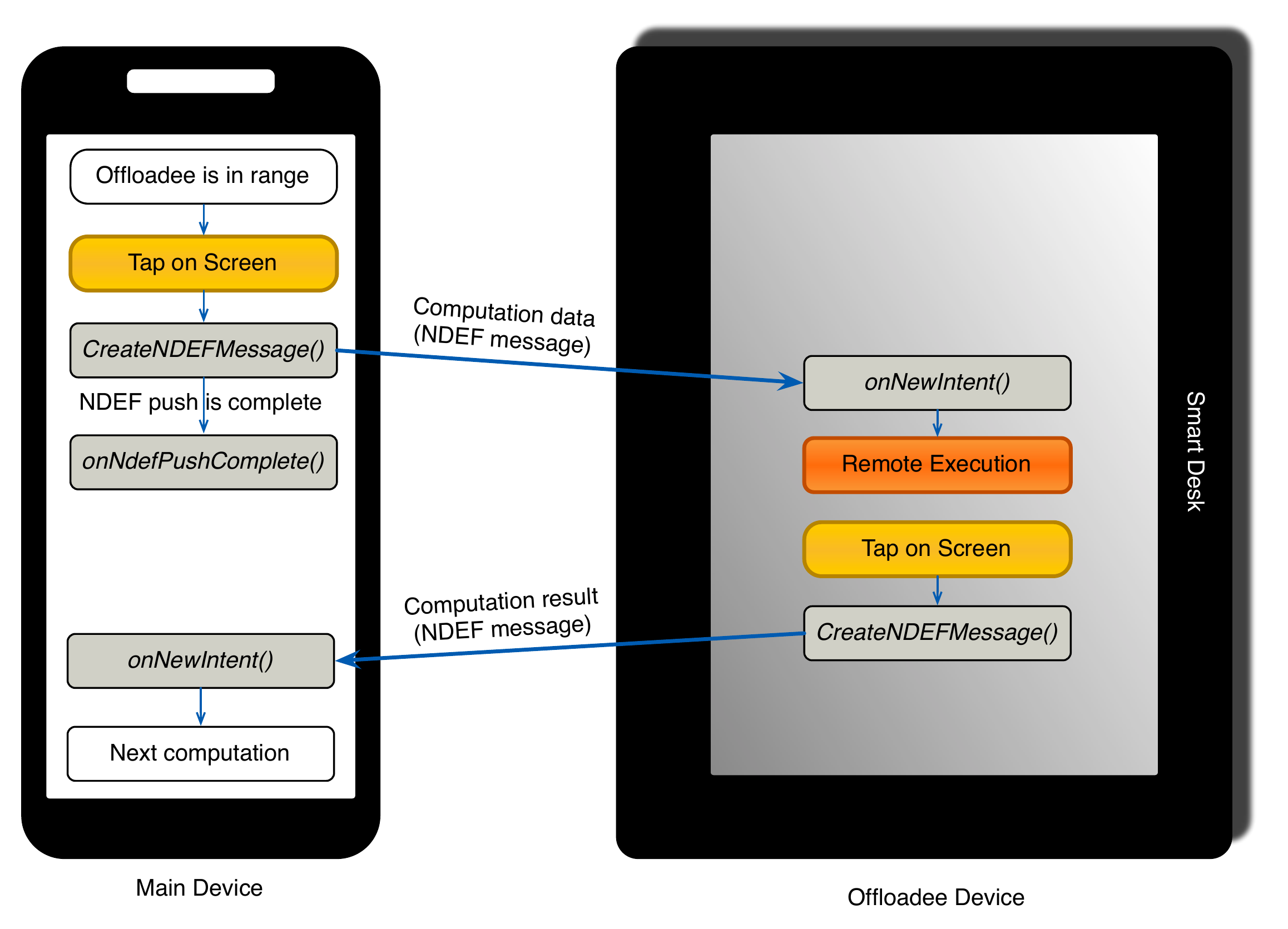}
    \caption{Two-tap Protocol. \vspace{-0.5cm}}
    \label{fig:twoTapImpl}
\end{figure}

We implemented a functional prototype which requires user intervention in several steps, as depicted in Figure \ref{fig:twoTapImpl}. Precisely, the user has to tap once on the main device's screen to send the task for remote computation to the offloadee device. Once the remote computation is finished, the user has to tap on the offloadee's screen to send back the result to the main device. The need for constant user intervention makes this strategy unusable in practice. In particular, even if the requirement of taping the main device---which usually be the user's device---could be tolerated, the requirement of the second tap on the offloadee device is not realistic. 
For this reason, we investigate other solutions that allow more flexibility and transparent implementation.

\subsubsection{Utilising Host-based Card Emulation}
Tapping only once at the beginning and allowing the offloadee device to automatically send back the result without a second tap creates more convenient user experience and can work on surface devices. In order to implement such functionality, we use the Android NFC Host-based Card Emulation (HCE) service\footnote{\label{ft:hce}\url{http://developer.android.com/guide/topics/connectivity/nfc/hce.html}}, which allows any NFC-enabled smartphone to emulate an NFC card so that it can be read directly by an NFC card reader.
In our case, the offloadee device emulates the NFC card while the main device emulates the NFC card reader. The card reader communicates with the emulated card by exchanging application-level packets called Application Protocol Data Units (APDUs) and by means of Application ID (AID), which are used as application selectors. 

\begin{figure}[t]
\centering
    \includegraphics[width=0.75\columnwidth]{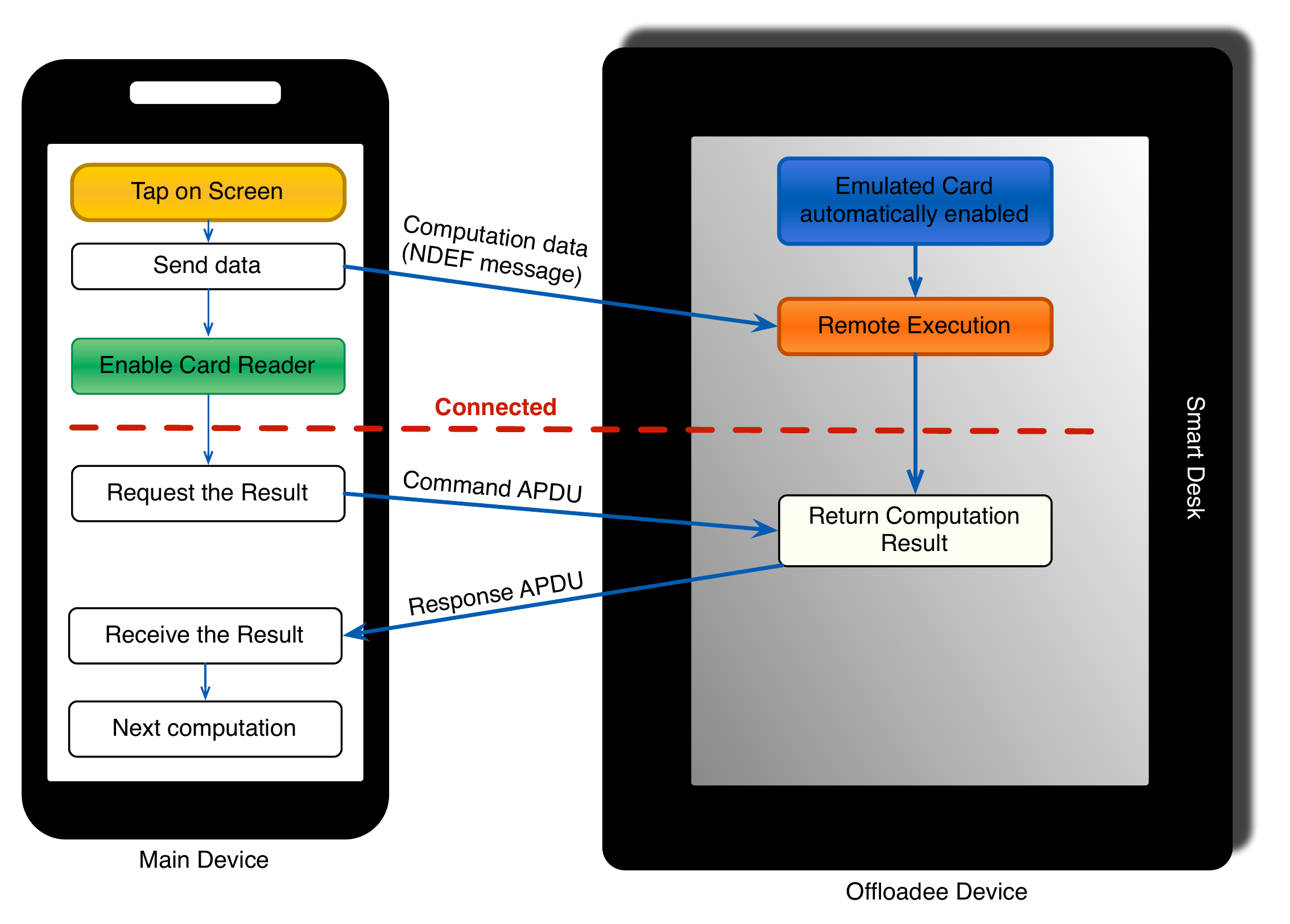}
    \caption{HCE Protocol. \vspace{-0.5cm}}
    \label{fig:hceImpl}
\end{figure}

When the environment is set up, the card reader searches for the emulated card and once the emulated card is detected the the card reader creates an APDU command to be sent to the emulated card to read the desired data. The command consists of a header and an AID. Once the emulated card receives the command, if the AID specified in the command is the same as the AID of the application, the function can immediately return the desired data concatenated with a few bytes designating the status word, which specifies that it is the message in response to the initial APDU command.Upon receiving the response, the card reader checks the status word bytes to ensure that it is the desired message.

The HCE Implementation is depicted in Figure~\ref{fig:hceImpl}. The most significant advantage of using the HCE implementation compared to the basic NdefMessage version is that it allows the main device and the offloadee device to communicate without tapping.
However, it is difficult to implement for an offloading framework, since the current Android NFC API allows only one role for each device, either as an NFC reader or as an emulated card. The solution we adopt in this implementation was to first utilise the basic \textit{NdefMessage} transfer method to send computation data before utilising the HCE service. When the user taps on the screen, the main device sends the computation data in NDEF format. Once the transfer is complete, the main device is transformed into a card reader. The offloadee, on the other hand, once receives and executes the computation, is transformed into an emulated card. The main device will then receive the computation result(s) automatically by reading the emulated card on the offloadee device. The limitations of this implementation are twofold: \textit{(i)} the user still needs to tap the screen on the main device and \textit{(ii)}  the offloading process can be realised only once, since the main device becomes a card reader and is not able to send NDEF messages anymore.

\subsection{Towards No-tap, Multiple Transfer Offloading}
We enabled multiple data exchange between the two devices by enabling both the card reader function and the card emulation function alternately on each smartphone.
This approach requires both devices to constantly switch roles until all the computation offloading is completed, which justifies that the implementation is not trivial.
However, we explore and implement two different strategies, namely: \textbf{(i)} the reader mode \textbf{disabling-enabling} method and \textbf{(ii)} the reader mode \textbf{enabling-disabling} method. Other than the number of transfers, there are two other significant advantages these methods possess:

 \textbf{1) No tapping is required}. Both methods only utilize the HCE service and therefore, data reading works automatically without any tapping.

 \textbf{2) Only one identical application is required}. While other methods require different applications to be installed on the main and offloadee devices---i.e. client--server components, this method only requires the same identical application to be installed on both devices.

\begin{figure}[t]
\centering
    \includegraphics[width=0.85\columnwidth]{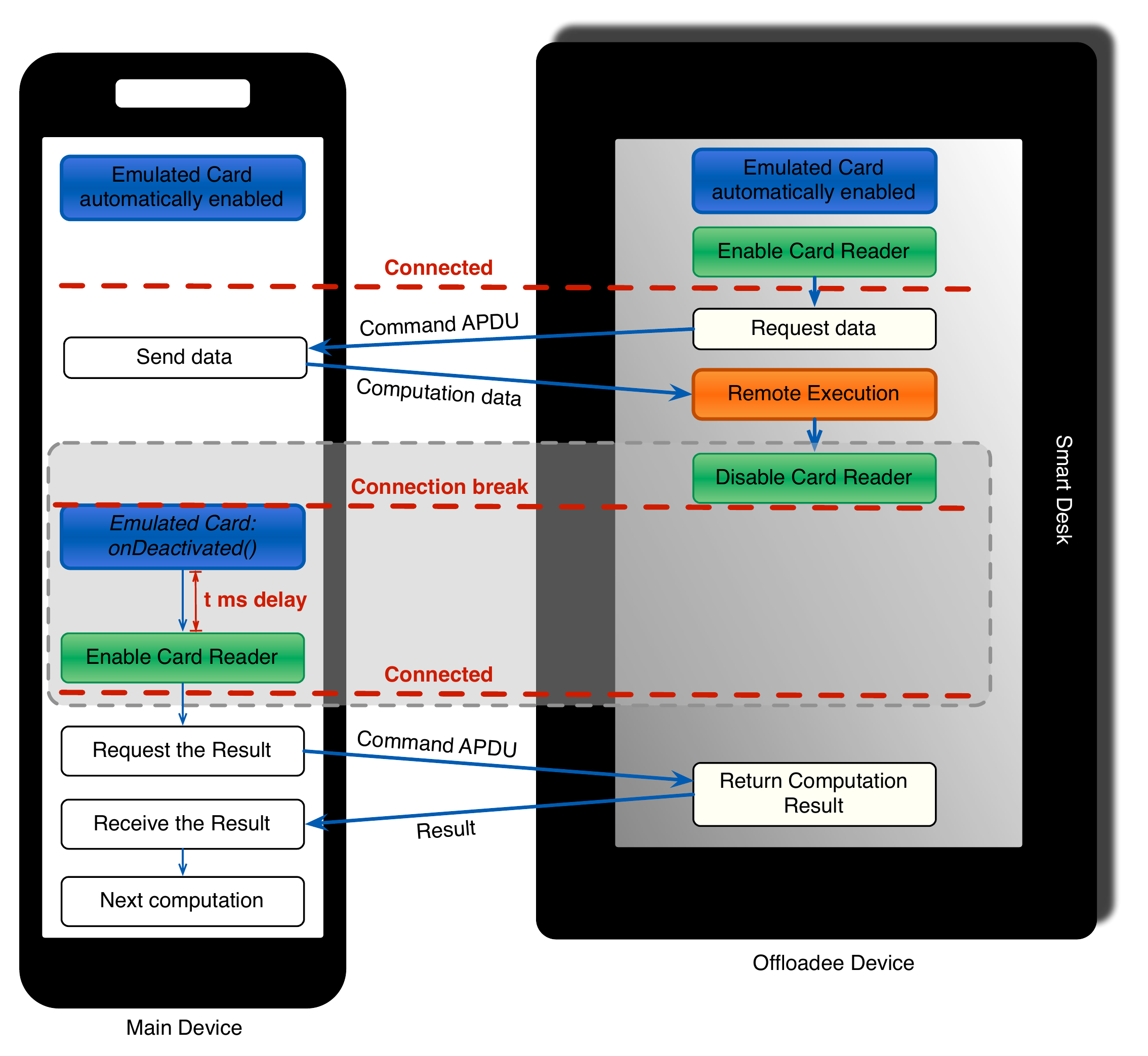}
    \caption{The Reader Mode \textbf{Disabling-Enabling} Protocol \vspace{-0.5cm}}
    \label{fig:DEA}
\end{figure}

\subsection{The Reader Mode \textbf{Disabling-Enabling}}

As summarized in Figure~\ref{fig:DEA}, the main idea of this method is to disable the reader mode on the one device before enabling the reader mode on the other device .
The implementation relies on the emulated card service method called \textit{onDeactivated()}, which will be called only when the connection to the card reader is lost. As mentioned before, the CardEmulationService allows the emulated card mode to be automatically enabled. In order to switch role to the card reader mode, we found that the device only needs to call the \textit{enableReaderMode()} function. Similarly, calling \textit{disableReaderMode()} would switch the role back to the emulated card mode.

At the beginning, the main device acts as an emulated card, while the offloadee device acts as a card reader. When the offloadee device finishes executing the offloadable task, it immediately switches role and becomes an emulated card by disabling the card reader mode. Once the card reader mode is disabled, the connection link to the emulated card on the main device breaks down. This triggers the \textit{onDeactivated()}, which then alerts the main device to switch role and become a card reader so that it can read the computation result from the offloadee device. 

There are two challenges in the implementation of this strategy: \textit{(i)} enabling role switching and \textit{(ii)} solving the hardware delay problem.
The first one is the trickiest. Even though we only need to call \textit{enableReaderMode()} and \textit{disableReaderMode()}, those functions can only be called from an Activity or a FragmentActivity class. The solution we adapt is to create a central Activity class, namely \textit{CentralActivity}. 
In terms of switching role from the card reader to the emulated card, we apply the procedure utilised by Android's CardReader sample application. Using this procedure, the lines of code listed below are added into the card reader class. 

\begin{lstlisting}[frame=single,language=java]
private WeakReference<MessageCallback> messageCallBack;
public interface MessageCallback(){
	public void onMessageReceived();
}
public CardReader (MessageCallback msg){
this.messageCallBack=new WeakReference<MessageCallback>(msg);
}
\end{lstlisting}

Then, the CentralActivity class is modified to implement \textit{CardReader.MessageCallback} and a new override method \textit{onMessageReceived()} is added. We finally put \textit{disableReaderMode()} within the override method. By applying this procedure, every time the card reader finishes interpreting the received message, it only needs to call \textit{messageCallback.get().onMessageReceived()} to disable the reader mode. In order to switch role from the emulated card to the card reader, once \textit{onDeactivated()} is called, a new intent is created to start CentralActivity. Once the activity runs, it will execute \textit{enableReaderMode()} within the \textit{onNewIntent()} method.

Regarding the second challenge, during the testing phase we notice that if we directly enable the reader mode once the \textit{onDeactivated()} is called, the new connection will not be created, and the card reader will not be able to read the emulated card. 
We then discover that the hardware needs some small amount of time before it can be ready to enable the reader mode. The process of selecting the proper amount of required delay is discussed in Section~\ref{sec:delay}.

\begin{figure}[t]
\centering
    \includegraphics[width=0.85\columnwidth]{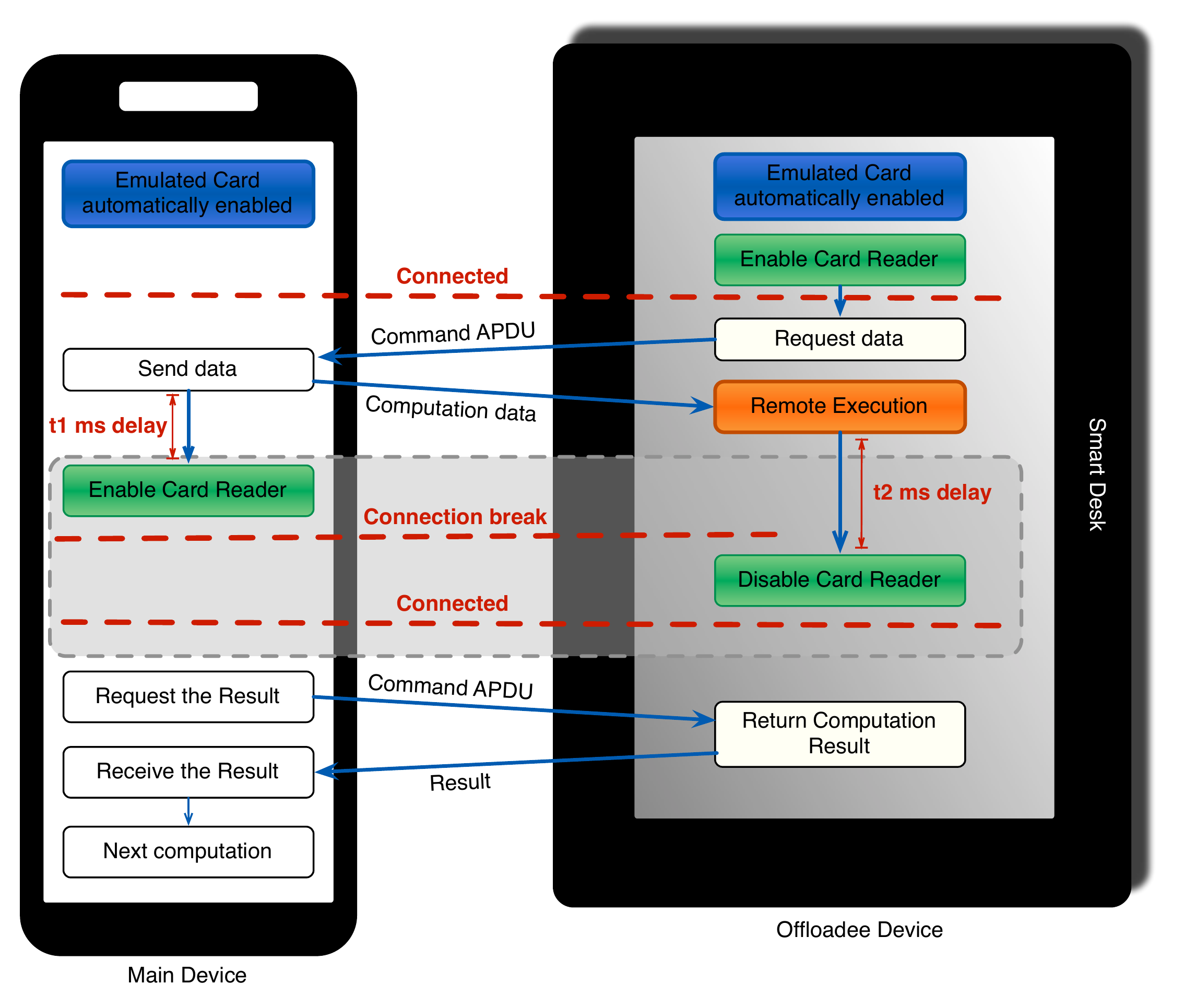}
    \caption{The Reader Mode \textbf{Enabling-Disabling} Protocol \vspace{-0.5cm}}
    \label{fig:EDA}
\end{figure}

\subsection{The Reader Mode \textbf{Enabling-Disabling}}

This implementation differs from the previous one only in the order we disable and enable the reader mode on each device.
By making such change in the implementation protocol, we discover that if we manage to enable the reader mode on one device before disabling the reader mode on the other, no delay is required; the connection immediately starts once the disabling reader mode occurs. The summary of this implementation is given in Figure~\ref{fig:EDA}.

We need to enable the card reader on the main device while the connection is running. One solution is to create a new thread which initializes a new intent for starting the central activity. One important note, however, is to ensure that the intent starts after \textit{processCommandApdu()} has returned the APDU response. Therefore, we set a delay on this new thread to start after $t_1$ ms. 
On the offloadee device, we have to ensure that the reader mode is disabled after the card reader on the main device has been enabled. Again, we set a delay on the disabling card reader of $t_2$ ms. We perform extensive experiments to find the appropriate values for $t_1$ and $t_2$. We discuss these findings in Section~\ref{sec:delay2}.

\section{Choosing the most performant NFC protocol}\label{sec:eval}
In this section we present the experiments we perform to measure the performances of the reader mode \textit{disabling-enabling} and \textit{enabling-disabling} protocols in terms of latency and bandwidth.
We evaluate them on two Xiaomi Mi~3 and one Samsung Galaxy Note~3 phones, whose specifications are in Table~\ref{tab:devices}.
The basic experiment setup consists on using one of the devices as the \textit{main device} and another  as the \textit{offloadee}.

\begin{figure*}[t]
\centering
\begin{subfigure}{0.65\columnwidth}
\centering
    \includegraphics[height=\columnwidth , angle=270]{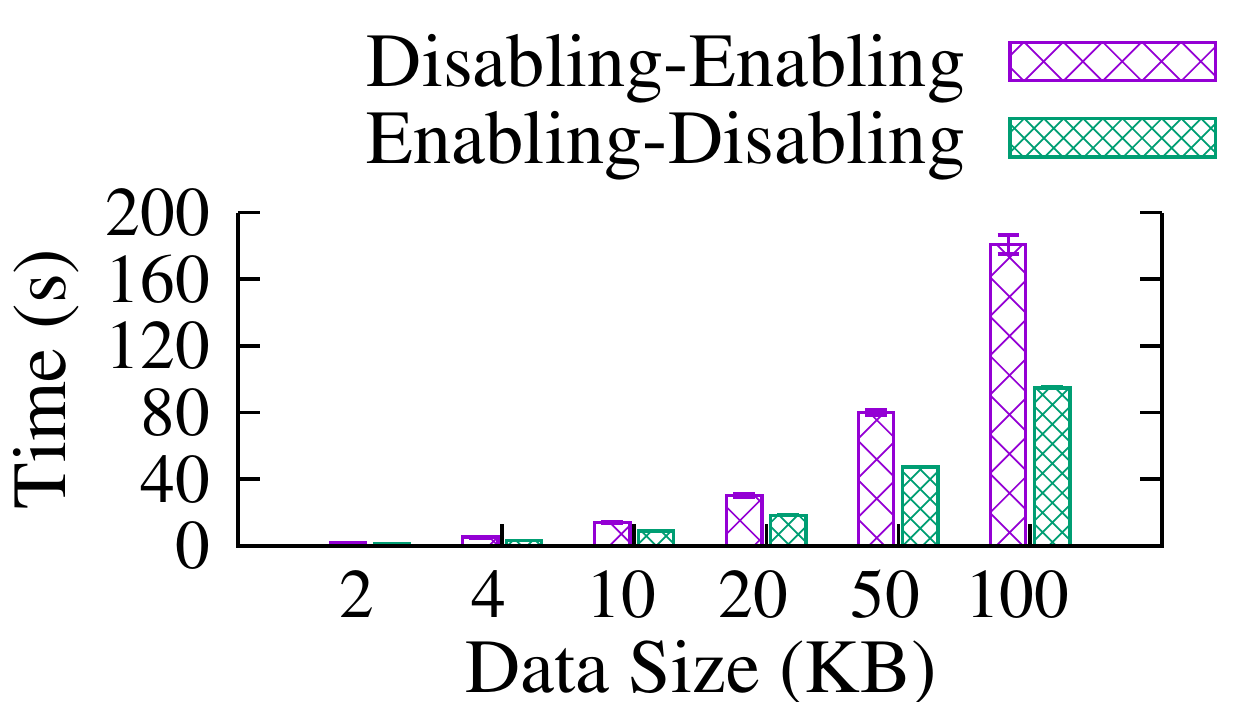}
    \caption{Latency}
    \label{fig:ppTime}
\end{subfigure}
\begin{subfigure}{0.65\columnwidth}
\centering
    \includegraphics[height=\columnwidth, angle=270]{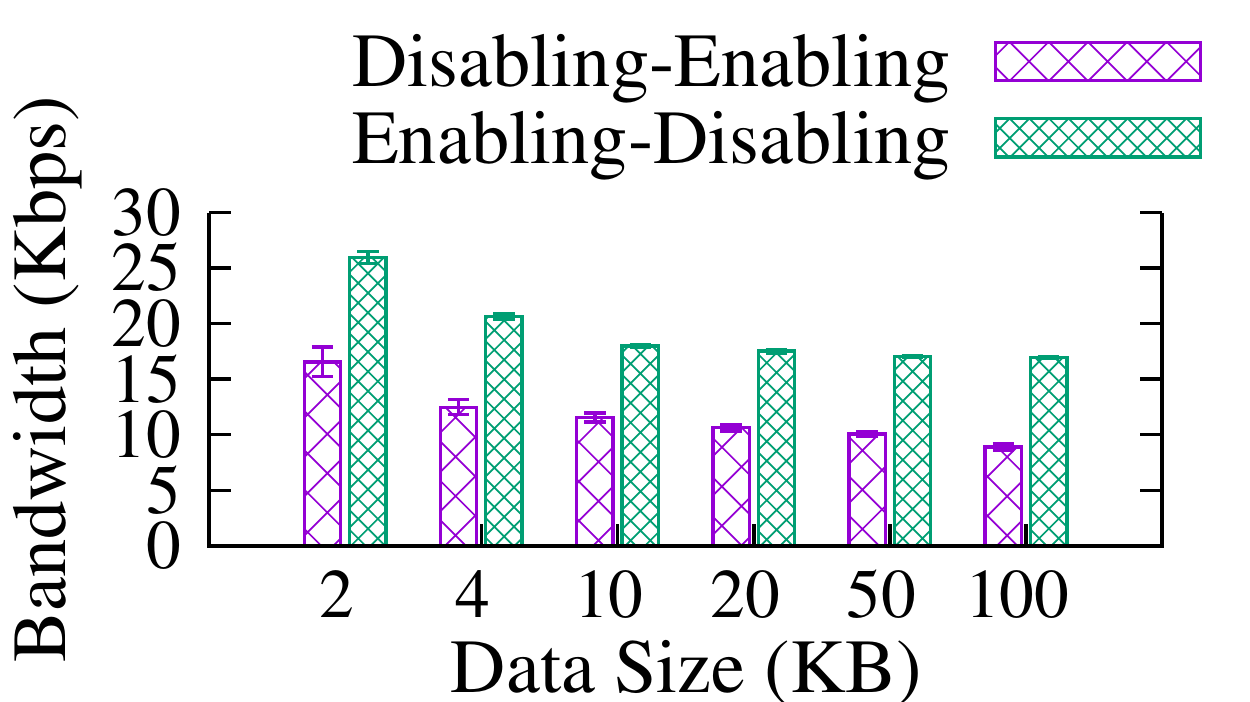}
    \caption{Bandwidth}
    \label{fig:ppBandwidth}
\end{subfigure}
\begin{subfigure}{0.65\columnwidth}
\centering
    \includegraphics[height=\columnwidth, angle=270]{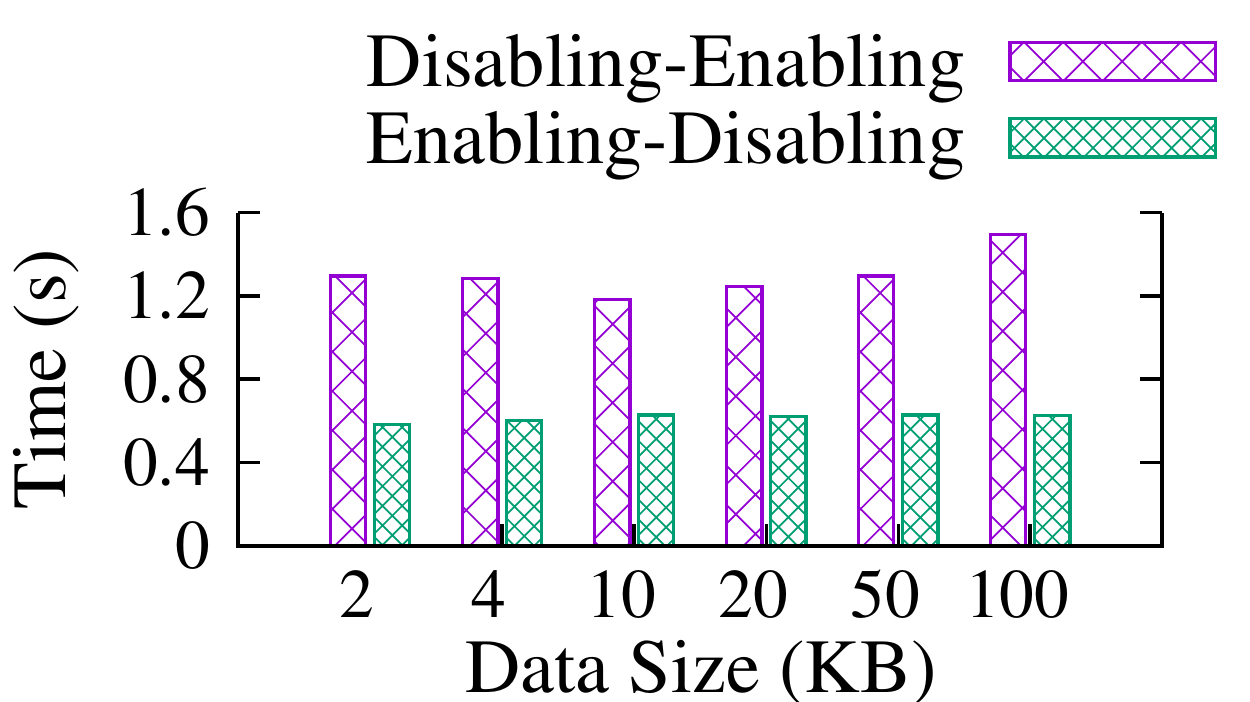}
    \caption{Role Switching Delay}
    \label{fig:ppProcTime}
\end{subfigure}
\caption{Latency, Bandwidth, and Role Switching delay when sending data of different size using the \textit{Disabling-Enabling} protocol with delay $t=700$ ms and the \textit{Enabling-Disabling} protocol with delays $t_1 = 310$ ms and $t_2 = 100$ ms. \vspace{-0.3cm}}
\label{fig:tbp}
\end{figure*}

\subsection{Testing the Reader Mode \textbf{Disabling--Enabling} Protocol}\label{sec:delay}
As we can see from Figure~\ref{fig:DEA}, the most important parameter of the \textit{disabling--enabling} protocol is the delay $t$.
Small values of $t$ could enable low latency and high bandwidth, but could increase the chances of transmission failure, since the hardware may not be able to switch in such short time.

We conduct an experiment using different values of $t$, while sending and receiving messages of 2KB for 50 times (round--trips).
Notice that we perform this round--trip experiment of small data, instead of single transmission of larger data, due to the limitations of the NFC packets.
Indeed, Android guidelines recommend the NFC messages to be smaller than 1KB. %\footref{ft:hce} 
but our experiments show that these packets can contain up to 2KB of data.
So, if an application wants to transmit more than 2KB it has to do so by performing more than one round--trip.
\begin{wraptable}[10]{l}{4.7cm}
 \vspace{-0.2cm}
 \centering
    \begin{tabular}{c c}
    \toprule
    $t$  (ms)  &  Success Rate    \\
	\cmidrule(r){1-2}
	680 & 5$\%$  \\  
	690 & 40$\%$  \\ 
	700 & 82$\%$  \\ 
	710 & 82$\%$  \\ 
	\bottomrule 
	\end{tabular}
   	\caption{The success rate of the \textbf{disabling-enabling} protocol for variable values of $t$.}
	\label{tab:tab2}
\end{wraptable} 
This process is handled automatically by our offloading framework, as we explain in detail in Section~\ref{sec:api}.
We consider the experiment successful if and only if \textbf{all} 50 round trips were correctly performed.
We repeat the experiment 20 times for each value of $t$ and count the number of successful experiments, which divided by 20 gives the success rate: percentage of experiments that successfully accomplished 50 round--trips.
The results of these experiments are presented in Table~\ref{tab:tab2}, from which we select the smallest value of $t$ such that the success rate is at least 80$\%$, which is $t = 700$ ms.

\subsection{Testing the Reader Mode \textbf{Enabling--Disabling} Protocol}\label{sec:delay2}
As we can see from Figure~\ref{fig:EDA}, the \textit{enabling--disabling} protocol is characterized by \textbf{two} delays: $t_1$, which is needed to enable the card reader, and $t_2$, which is need to disable the card reader mode.
One indicator for finding $t_1$ is to look for the lowest possible time delay while ensuring that the role switching occurs after the device has sent an APDU response.
\begin{wraptable}[15]{l}{5cm}
 \vspace{-0.2cm}
 \centering
\begin{tabular}{c c}
    \toprule
    $t_1$  (ms)  &  Success Rate    \\
	\cmidrule(r){1-2}
	250 & 0$\%$  \\  
	260 & 0$\%$  \\
	270 & 30$\%$  \\
	280 & 55$\%$  \\
	290 & 60$\%$  \\
	300 & 65$\%$  \\
	310 & 95$\%$  \\
	\bottomrule 
	\end{tabular}
   	\caption{The success rate of the \textbf{enabling-disabling} protocol for delay values $t_2=1000$ ms and variable $t_1$.}
	\label{tab:tab3}
\end{wraptable} 
In searching for the optimal $t_2$, on the other hand, we have to make sure that the reader mode is disabled after the reader mode on the other device is enabled.
When at least one of these values is too low, the round--trip data transmission will stop with error message: ``\textit{Error communicating with card: android.nfc.TagLostException: Tag was lost}".We first find the value of $t_1$ by initially setting $t_2$ equal to 1000 ms so that it will not hinder the round--trip transmission.
\begin{wraptable}[11]{l}{5cm}
\vspace{-0.2cm}
 \centering
    \begin{tabular}{c c}
    \toprule
    $t_2$  (ms)  &  Success Rate    \\
	\cmidrule(r){1-2}
	50 & 0$\%$  \\  
	70 & 0$\%$  \\ 
	90 & 0$\%$  \\ 
	100 & 85$\%$  \\ 
	\bottomrule 
	\end{tabular}
   	\caption{The success rate of the \textbf{enabling-disabling} protocol for $t_1 = 310$ ms and variable $t_2$. \vspace{-0.5cm}}
	\label{tab:tab4}
\end{wraptable} 

Once $t_1$ is selected, $t_2$ can then be determined similarly. We follow the same process as in the previous section, sending round--trip messages of 2 KB for 50 times, to measure the success rate of the experiments.
The results are presented in Table~\ref{tab:tab3} and Table~\ref{tab:tab4}. From the first experiment we fixed $t_1=310$ ms and from the second $t_2=100$ ms.

\subsection{\textbf{Disabling--Enabling} vs. \textbf{Enabling--Disabling} Protocol}\label{sec:comparison}
Figures~\ref{fig:ppTime} and ~\ref{fig:ppBandwidth} show the latency and the bandwidth results of the two protocols when sending data of different size.
The used delay values are $t=700$ ms for the \textit{disabling--enabling} and $t_1 = 310$ ms, $t_2 = 100$ ms for the \textit{enabling--disabling}.
The \textit{enabling--disabling} protocol presents better performance, with latency being around $50\%$ smaller and bandwidth being about $1.6$ times higher compared to the \textit{disabling--enabling}.

\begin{figure}[t]
\centering
    \includegraphics[width=.6\columnwidth]{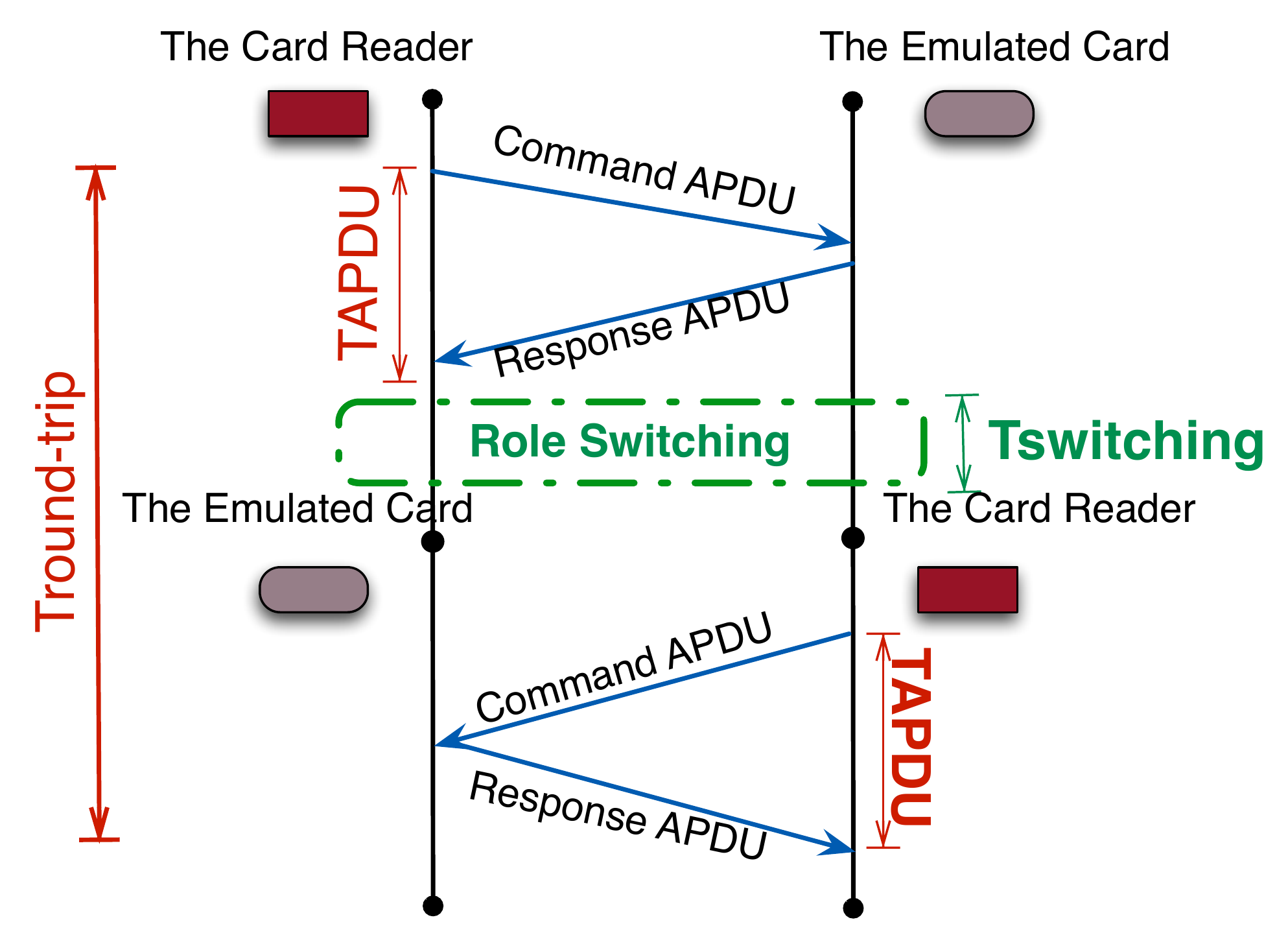}
\caption{Definition of $T_{APDU}$, $T_{switching}$, and $T_{round-trip}$. \vspace{-0.3cm}}
\label{fig:plots3233}
\end{figure}

To better understand and calculate the duration of the role--switching process, we show in Figure~\ref{fig:plots3233} our NFC protocol of sending the data in round--trip.
We define $T_{APDU}$ as the time needed to send an APDU command from the card reader to the emulated card and sending the APDU response backto the card reader.
We define $T_{switching}$ as the time needed for both devices to switch roles.
Finally, we define $T_{round-trip}$ as the total time it takes for one device to send the request, for devices to switch roles, and for the device to get the response back.
As we can see, the formula to calculate the time for one round--trip can be expressed by the following equation: 
\begin{equation*}
	T_{round-trip}^{(1)} = 2 \cdot T_{APDU} + T_{switching}.
\end{equation*}
The formula for 2 round--trip transmissions, assuming $T_{switchingAvg}$ is the average of all $T_{switching}$, is:
\begin{eqnarray*}
T_{round-trip}^{(2)}&=& 2 \cdot T_{round-trip}^{(1)} + T_{switchingAvg} \nonumber \\
	&=& 4 \cdot T_{APDU} + 3 \cdot T_{switchingAvg}.
\end{eqnarray*}
Iterating the formula for $n$ round--trips, we obtain:
\begin{eqnarray*}
	T_{round-trip}^{(n)}= 2n\cdot  T_{APDU} + (2n-1)\cdot  T_{switchingAvg} \\
\text{Therefore: }	T_{switchingAvg} = \frac{T_{round-trip}^{(n)} - 2n\cdot  T_{APDU}}{2n - 1}.
\end{eqnarray*}
Knowing, from experimental results, that $T_{APDU}$ for a message of 2 KB is 329 ms, we can use the previous formula to calculate the average value of the switching time for both protocols.
In Figure~\ref{fig:ppProcTime} we show the calculated values of $T_{switchingAvg}$ when sending data of different size.
Based on these results, it is apparent that the average switching time of the reader mode \textit{enabling--disabling} protocol is less than half of the reader mode \textit{disabling--enabling} protocol.

\section{Final Offloading Framework}\label{sec:univ} 

After extensive testing and evaluation, presented in Section~\ref{sec:eval}, we conclude that the \textbf{enabling-disabling} strategy presents the \textit{lowest data transmission delay} and \textit{highest bandwidth} of all proposed protocols. 
Hence, we build the offloading framework library on top of this communication protocol.

\subsection{Supported API}\label{sec:api}

To enable bidirectional transmission of large quantity of data, we design a \textit{MessageStorage} class which stores two two-dimensional byte arrays of \textit{messageToSend}--the message which will be sent by the emulated card as an APDU response, and \textit{messageReceived}--the message which is received by the card reader from the emulated card.
This class implements a data transmission protocol based on the \textit{enable--disable} protocol, which allows developers to transparently send and receive large quantity of data that would be conversely impossible to achieve with the default NFC protocol.
The class exposes the following methods:

\textbf{$\bullet$ \mbox{SetMessageToSend(byte[] message, int index)}} sets the $messageToSend[index]$ to the message value. It is called by the emulated card to prepare the APDU response.

\textbf{$\bullet$ \mbox{GetMessageToSend(int index)}} returns the value of $messageToSend[index]$. It is called by the emulated card upon receiving an APDU command. The emulated card then sets it as the APDU response and sends it to the reader.

\textbf{$\bullet$ \mbox{SetMessageReceived(byte[] message, int index)}} sets the $messageReceived[index]$ to the message parameter value. Once the card reader receives an APDU response from the emulated card, it stores the value by calling this method.

\textbf{$\bullet$ \mbox{GetMessageReceived(int index)}} returns the value of $messageReceived[index]$. It is called by the card reader to retrieve the result received by the emulated card.

On the emulated card, the APDU response is constructed by storing the desired message into the message byte array and calling the \textit{SetMessageToSend(message, 0)} function. If the message is bigger than 2KB, we divide it into $n$ arrays of size 2KB or smaller. Then, each of these arrays is stored sequentially inside the class object.
When the emulated card receives the command APDU containing the AID, it reads the last 2 digits to get the index and returns \textit{GetMessageToSend(index)}. Once the card reader receives the message, it immediately calls \textit{SetMessageReceived(received$\_$message, index)}. For further processing, it is able to get the received message easily by calling \textit{GetMessageReceived(index)}.

\begin{figure*}[t]
\centering
\begin{subfigure}{\columnwidth}
\centering
    \includegraphics[width=0.66\columnwidth]{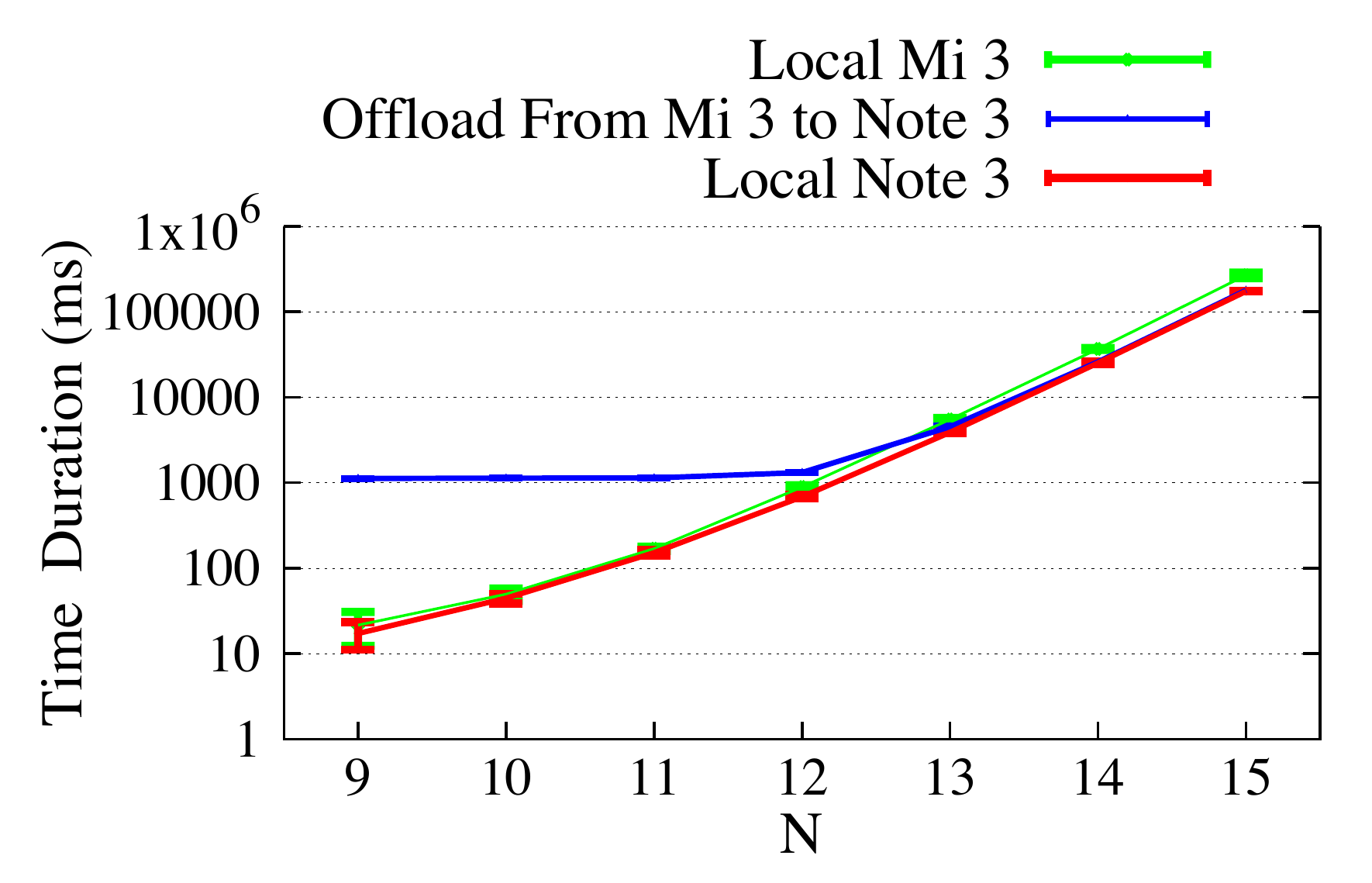}
    \caption{The N Queens puzzle execution time.}
    \label{fig:NQtime}
\end{subfigure}
\begin{subfigure}{\columnwidth}
\centering
    \includegraphics[height=0.66\columnwidth, angle=270]{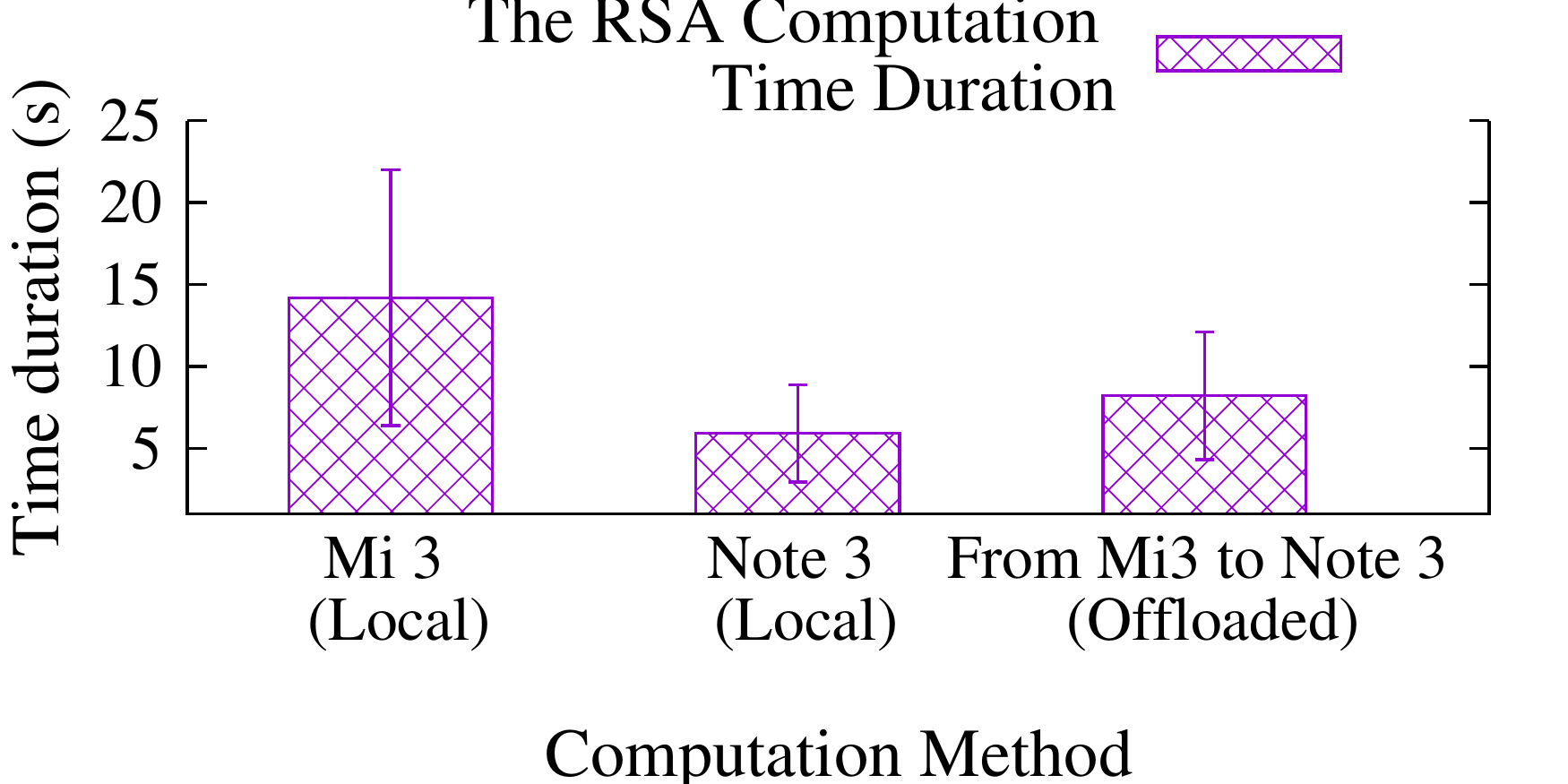}
    \caption{The RSA execution time.}
    \label{fig:RSAtime}
\end{subfigure}
\begin{subfigure}{0.65\columnwidth}
\centering
    \includegraphics[height=\columnwidth , angle=270]{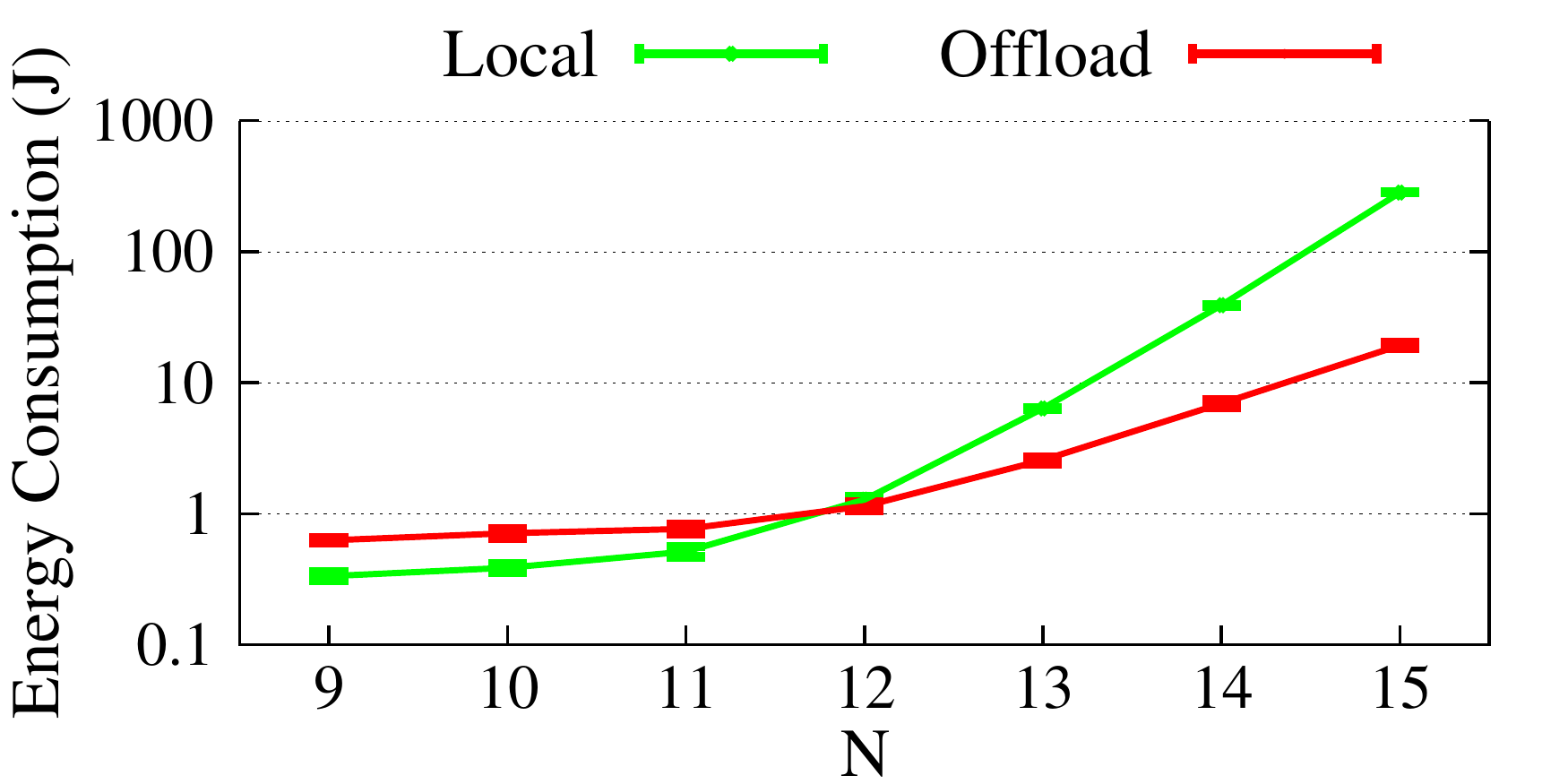}
    \caption{The N Queens puzzle energy consumption.}
    \label{fig:NQenergy}
\end{subfigure}
\begin{subfigure}{0.65\columnwidth}
\centering
    \includegraphics[width=\columnwidth]{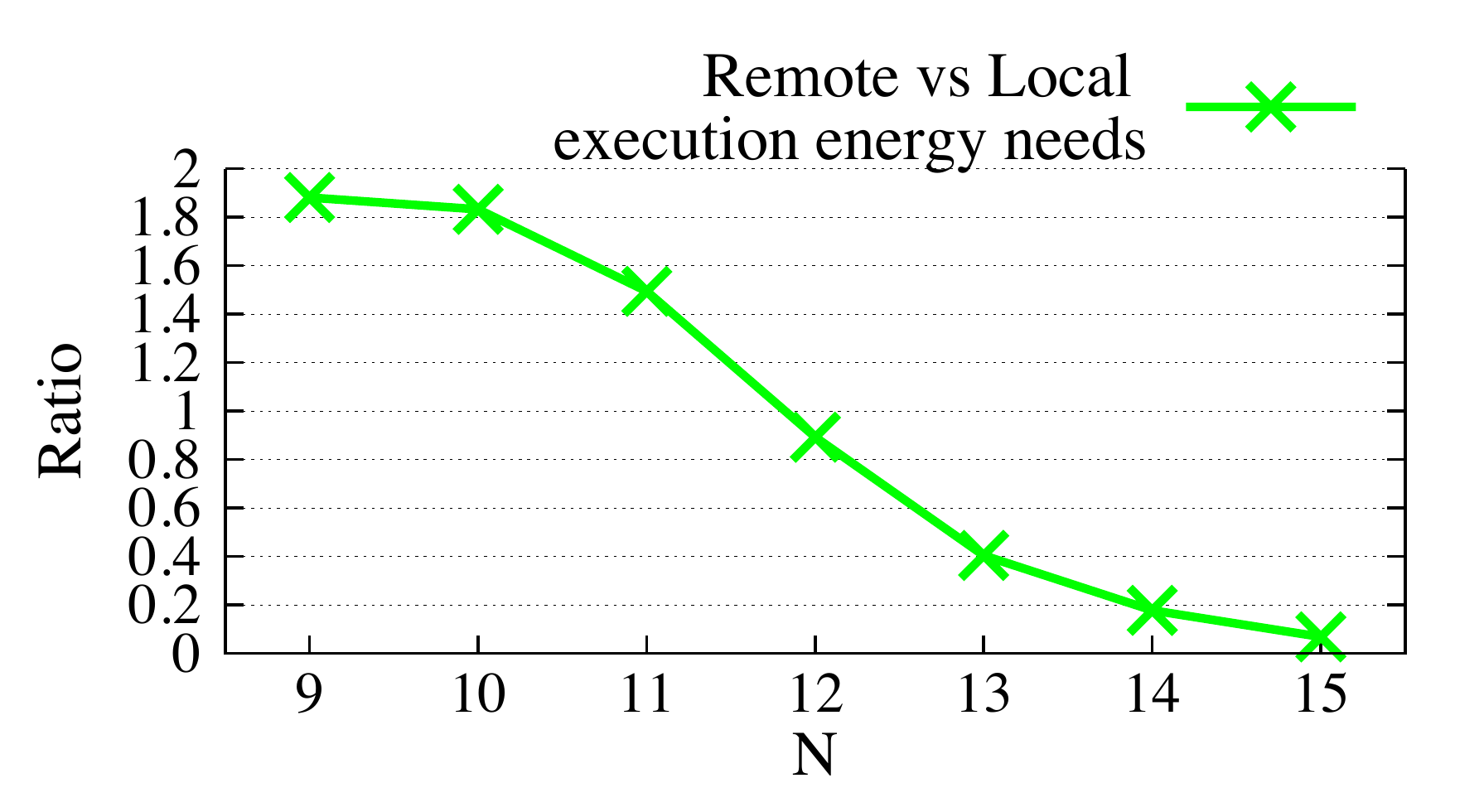}
    \caption{The local execution energy consumption by the offloaded execution energy consumption ratio of the N Queens puzzle.}
    \label{fig:NQenergyRatio}
\end{subfigure}
\begin{subfigure}{0.65\columnwidth}
\centering
    \includegraphics[height=\columnwidth, angle=270]{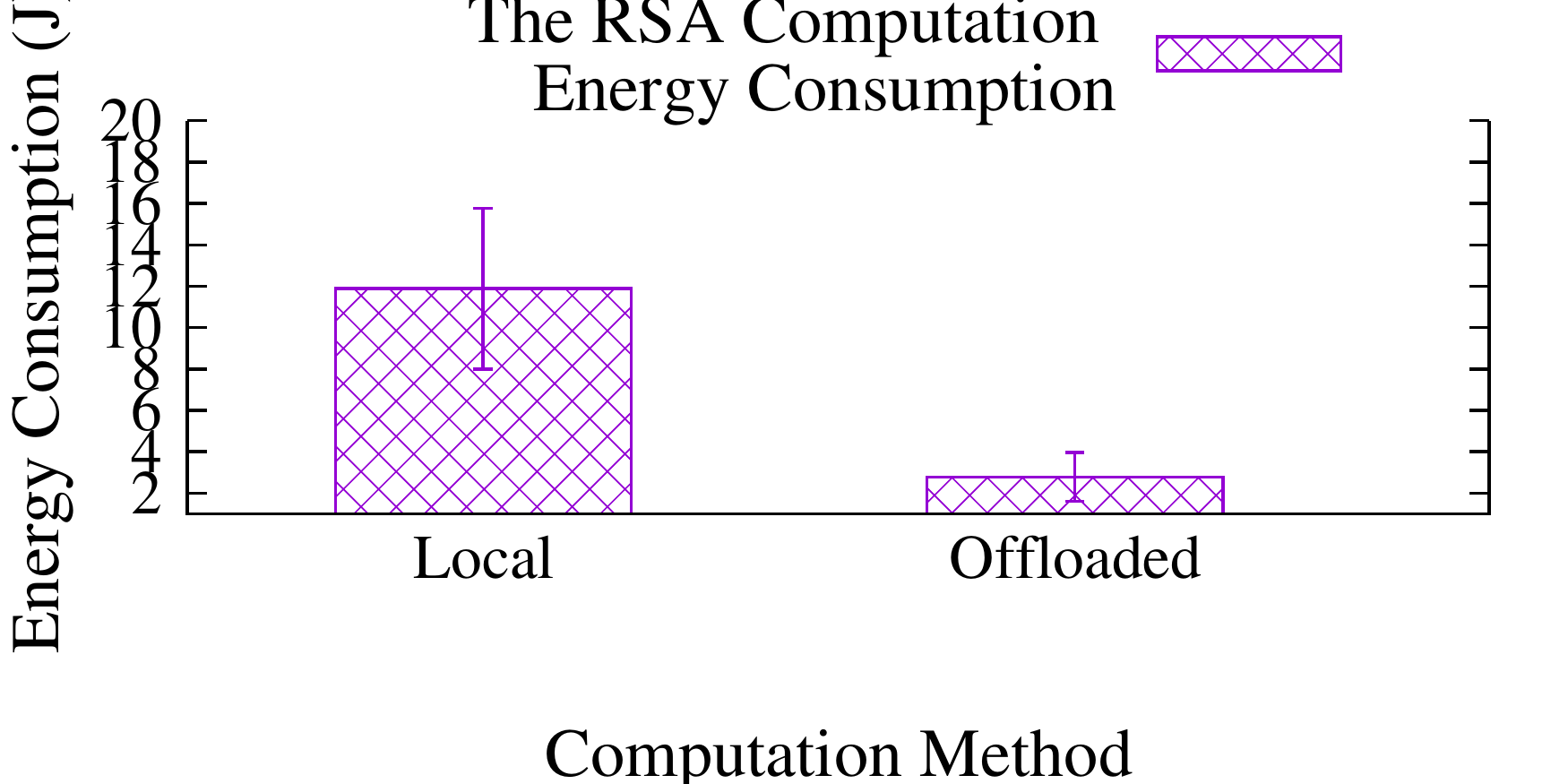}
    \caption{The RSA energy consumption.}
    \label{fig:RSAenergy}
\end{subfigure}
\caption{The execution time duration and the energy consumption of the N Queens puzzle and the RSA encryption algorithm. All the plots depict the average of 50 repetitions and the error-bars show the standard deviation.   \vspace{-0.3cm}} %
\label{fig:time}
\end{figure*}

\subsection{Advantages and Limitations}\label{sec:prosAndcons}

The advantages of our proposal can be categorised in the following four main points:
\textbf{(1) Fully automatic:} Our framework removes the requirement of tapping. We implement the new NFC protocol using the NFC/HCE service which doesn't require user intervention, making it possible to transparently run applications in a convenient user experience.
\textbf{(2) Portability:} Application developers do not need to implement specific application versions for different device roles, i.e. \textit{offloader} and \textit{offloadee}. They only need to install the same identical application on both devices and specify the role of each device through the framework's settings.
\textbf{(3) Application execution time improvement:} When a device offloads the heavy tasks of an application to a more powerful offloadee, the overall execution time of the application is reduced.
\textbf{(4) Device energy reduction:} The energy consumption of the device that offloads the heavy tasks is reduced, since it is the offloadee that takes care of the computation and because the NFC data transmission is very low--energy consuming.

The limitations of our framework, which are inherited by the existing underlying technologies are twofold: \textbf{(1)} Limited bandwidth and \textbf{(2)} small APDU packet size.
The main drawback that comes from these limitations is that the current implementation of the framework is not suitable for data--intensive applications, which would need to transfer high quantity of data during the offloading process.

\subsection{Characteristics of applications suitable for NFC offloading}\label{sec:apps}

Considering the pros and cons of our NFC protocol and of our NFC offloading framework, the best application candidates suitable for NFC offloading should have the following characteristics:
\textbf{(1)} small input size, \textbf{(2)} small output size, and \textbf{(3)} high computational needs.
Based on these, the proposed framework is appropriate for computationally intensive applications that do not require high data transfers, such as \textbf{(1)} encryption, \textbf{(2)} mobile payments, \textbf{(3)} cryptocurrencies, \textbf{(4)} mathematical computations, etc.

\section{Experiments}\label{sec:exp}

In order to evaluate the performance of our proposal we utilise two of the three puzzles that are proposed by Google in its Google Optimization Tools\footnote{https://developers.google.com/optimization/puzzles}. Specifically, we implement the N Queens mathematical puzzle~\cite{NQ} and the Rivest-Shamir-Adleman (RSA) encryption algorithm~\cite{RSA}. In the rest of this section, we initially present these two applications and then we discuss their performance in terms of \textbf{(1)} execution duration and \textbf{(2)} energy needs. 

\textbf{$\bullet$ N Queens} is a classic puzzle of placing N queens on a N x N chess board so that no queen can attack another in one move. In our implementation, we find how many are valid solutions for a given N. By using a backtracking algorithm, our algorithm complexity is O(n!). N Queens is a representative application that is generally adopted for its high computational requirements. Our mechanism is able to assist more sophisticated applications and we hope to be used on the implementation of future killer applications in the highly active area of NFC.  
During the offloading process of the N Queens Computation the main device, which initially is the emulated card, runs the N Queens application and the number and the inputted N are stored in the \textit{messageToSend} two-dimensional byte arrays in [application$\_$number | N] format. When the offloadee device, which initially is the card reader, is within range, the main device sends the value of \textit{messageToSend} as the response APDU. Once the offloadee device receives the details of the sample application, it immediately executes the computation based on the received N and stores the result in its \textit{messageToSend} variable. After both devices have switched roles, the main device reads the result from the offloadee device. 

\textbf{$\bullet$ RSA} is an asymmetric cryptographic algorithm and is comprised of three main parts: 1) \textit{Key generation:} the key generation process aims to generate public and private keys from two large prime numbers. Each prime number is at least 2048 digits, which is considered to be secure based on 2014 technology \cite{RSA}. 2) \textit{Encryption:}  a given plain text is encrypted using the generated public key from the previous process.  3) \textit{Decryption:} the private key is used  to decrypt the encrypted text. We offload the key generation and encryption processes in this application. The decryption is performed on the main device after the offloading is finished to ensure that the transferred data are not corrupted. We use the java.security API with 2048 bits as the key length in the key generation process. Given a plain text as the input that is no more than 2048 bits, the computation produces a set of private and public keys and the encrypted message.
On the main device, the application prompts the user for the plain text file. Once the user presses the Start button, the application reads the file and gets the plain text in bytes. The plain text is then stored in \textit{messageToSend}. When the offloadee device is within range, the main device sends the message stored in \textit{messageToSend}. Upon receiving the message, the offloadee immediately starts the RSA computation: keys generation and encryption. It then stores the public key, the private key, and the decrypted text in \textit{messageToSend}. Since the total lengths of the result is more than 2 KB, it is divided into two separate byte arrays. The first byte array stores the concatenated decrypted text and the public key because the decrypted text is always 512 B (based on the key length) while the public key is always less than 1500 B. The second byte array stores the private key. After both devices switch roles, the main device reads the whole result by sending two separate APDU commands and store all the received responses in \textit{messageReceived}. 

\subsection{Execution Time}\label{sec:time}

In the case of N Queens, we measure the execution time for $ N = \{9,10,11,12,13,14,15\}$  using a Xiaomi Mi 3 and a Samsung Galaxy Note 3. 
We measure the time for the case of the local execution, in both devices, and for the remote execution we use the Xiaomi Mi 3 as the main device and the Samsung Galaxy Note 3 as the offloadee device. 
This is because based on the results of the local execution, the computations on the Samsung Galaxy Note 3 phone are much faster than the computation on the Xiaomi Mi 3,  therefore using the Samsung Galaxy Note 3 as the offloadee device is be more advantageous. 
Note that the duration we measure is the time the main device needs to offload the application and receive the output of the remote execution.  
For small values of $N$, the offloaded application has worse performance that the local ones regardless of the used device. This is due to the communication overhead. But for high values of $N$,  the Xiaomi Mi 3 requires much more time if it executes the application locally instead of offloading it. 
For the case of the RSA application, we measure the time duration of the local execution (on both devices) and the offloaded execution where the Xiaomi Mi 3 is the main device and the Samsung Galaxy Note 3 is the offloadee device. 
 Xiaomi Mi 3 requires almost 2.5 more time to execute the application locally than the Samsung Galaxy Note 3. However, if the application is offloaded, Xiaomi Mi 3 requires half time than it requires to execute it locally. 
Both plots \ref{fig:NQtime} and  \ref{fig:RSAtime} of Figure \ref{fig:time} show that the time required for the computation offloading is significantly close with the local computation duration of the offloadee device instead of the local computation duration of the offloader device. Hence, offloading computation using our framework on a more powerful offloadee device will provide shorter time duration than running it locally.

\subsection{Energy Consumption}\label{sec:energy}

We use the Samsung Galaxy Note 3 for the power measurement because in order to measure the energy consumption accurately we need to remove the battery and the battery of the Xiaomi Mi 3 is not removable. We use the highly adopted Monsoon Power Monitor\footnote{https://www.msoon.com/LabEquipment/PowerMonitor/}. 

For the N Queens application, we measure the energy consumption for $ N = \{9,10,11,12,13,14,15\}$ first for the case of the local execution
and then for the remote one. For the latter, we use the Samsung Galaxy Note 3 as the main device and the Xiaomi Mi 3 as the offloadee device because we are interested in the energy consumption of the main device. 
Figure \ref{fig:NQenergy} shows the results of both measurements. 
For small values of N (i.e. $N<12$) the offloading is not beneficial in terms of energy, but for $N\geq12$ the benefit is increasing and for the case of $N=15$, the main device consumes up to 15 times less energy due to the offloading. 
Figure \ref{fig:NQenergyRatio} shows how the ratio of the energy needs of the offloaded computation to the local computation in the N Queen problem is decreasing as the hardness of the computations is increasing. 

Regarding the RSA application we measure the energy consumption of both the local computation and offloaded computation of RSA key generation and encryption on a Samsung Galaxy Note 3. In the case of the remote execution, we use the Samsung Galaxy Note 3 as the main device and the Xiaomi Mi 3 as the offloadee device and we present the measurements on Figure \ref{fig:RSAenergy}.  
In the case of offloading, the main device requires less than 20$\%$ of the energy required by the case of local execution. 
By these experiments we argue that there is a significant benefit of using our framework to offload heavy computations.

\section{Conclusion and Future Work}\label{sec:concl}
In this paper we proposed and implemented the first NFC offloading framework for Android devices.
First, we proposed a new NFC communication protocol that circumvents the limitations of the default Android NFC protocol.
Our protocol eliminates the requirement of user intervention, working automatically without needing users to tap on the device's screen for data transfer.
Furthermore, our protocol enables a bidirectional communication between two devices, which paved the way towards building the NFC offloading framework.
Finally, we implemented the first known, to the best of our knowledge, NFC-based computation offloading framework between two smart devices.
We implemented two applications that use the framework to offload heavy computations from one \textit{main device} to an \textit{offloadee device}.
We showed that when the offloadee device is more powerful, the execution time of the offloaded task is improved.
We also showed that the main device is able to reduce its energy consumption when offloading the computations, due to the low--energy consumption of the NFC interface.

We observed several limitations of our framework, mostly due to the underlying technologies that NFC is built upon.
The main problem we faced was the low bandwidth of the data transmission, with values around $15-25$ Kbps, which is cause by the existing hardware that allows only one message per connection. 
These make the framework not suitable for several types of applications, in particular those that need to transfer many data during the offloading process. 
However, any new NFC chip that can support multiple messages per connection will increase the bandwidth significantly and broad the applicability of NFC. 
We plan to extend our framework to make it more heterogeneous by supporting other operating systems and devices, such as tablets, laptops, and desktops with external NFC readers.
Furthermore, we will progress the current API to expose more functionalities to the developers.
Finally, we are currently investigating techniques to increase the bandwidth and support a broader range of applications and to complement the NFC framework with parallel connections between the devices using Bluetooth and/or WiFi--direct. 

\section{Acknowledgements}
This research has been supported, in part, by General Research Fund 26211515 from the Research Grants Council of Hong Kong, Innovation and Technology Fund ITS/369/14FP from the Hong Kong Innovation and Technology Commission, and the European Commission under the Horizon 2020 Program through the RAPID project (H2020-ICT-644312).

%\eject
\bibliographystyle{IEEEtran}
\bibliography{Infocom17}

% Generated by IEEEtran.bst, version: 1.13 (2008/09/30)
\begin{thebibliography}{10}
\providecommand{\url}[1]{#1}
\csname url@samestyle\endcsname
\providecommand{\newblock}{\relax}
\providecommand{\bibinfo}[2]{#2}
\providecommand{\BIBentrySTDinterwordspacing}{\spaceskip=0pt\relax}
\providecommand{\BIBentryALTinterwordstretchfactor}{4}
\providecommand{\BIBentryALTinterwordspacing}{\spaceskip=\fontdimen2\font plus
\BIBentryALTinterwordstretchfactor\fontdimen3\font minus
  \fontdimen4\font\relax}
\providecommand{\BIBforeignlanguage}[2]{{%
\expandafter\ifx\csname l@#1\endcsname\relax
\typeout{** WARNING: IEEEtran.bst: No hyphenation pattern has been}%
\typeout{** loaded for the language `#1'. Using the pattern for}%
\typeout{** the default language instead.}%
\else
\language=\csname l@#1\endcsname
\fi
#2}}
\providecommand{\BIBdecl}{\relax}
\BIBdecl

\bibitem{cite1}
J.~Newman, ``{Peak Battery: Why Smartphone Battery Life Still Stinks, and Will
  for Years},''
  \url{http://techland.time.com/2013/04/01/peak-battery-why-smartphone-battery-life-still-stinks-and-will-for-years/},
  2013.

\bibitem{cite2}
S.~D. Ltd, ``{FlashBattery for smartphones},''
  \url{http://www.store-dot.com/#!smartphones/zoom/c1w5t/c1u51}, 2015.

\bibitem{cite3}
D.~Borghino, ``{Nanodot-based smartphone battery that recharges in 30
  seconds},''
  \url{http://www.gizmag.com/nanodot-smartphone-battery-30-second-recharge/31467/},
  2014.

\bibitem{Balan:2002:CCF:1133373.1133390}
R.~Balan, J.~Flinn, M.~Satyanarayanan, S.~Sinnamohideen, and H.-I. Yang, ``The
  case for cyber foraging,'' in \emph{Proceedings of the 10th Workshop on ACM
  SIGOPS European Workshop}, ser. EW 10, 2002, pp. 87--92.

\bibitem{MAUI}
E.~Cuervo, A.~Balasubramanian, D.-k. Cho, A.~Wolman, S.~Saroiu, R.~Chandra, and
  P.~Bahl, ``Maui: Making smartphones last longer with code offload,'' in
  \emph{Proceedings of the 8th International Conference on Mobile Systems,
  Applications, and Services}, ser. MobiSys '10, 2010, pp. 49--62.

\bibitem{CLONECLOUD}
B.-G. Chun, S.~Ihm, P.~Maniatis, M.~Naik, and A.~Patti, ``Clonecloud: Elastic
  execution between mobile device and cloud,'' in \emph{Proceedings of the
  Sixth Conference on Computer Systems}, ser. EuroSys '11, 2011, pp. 301--314.

\bibitem{kosta2012thinkair}
S.~Kosta, A.~Aucinas, P.~Hui, R.~Mortier, and X.~Zhang, ``Thinkair: Dynamic
  resource allocation and parallel execution in the cloud for mobile code
  offloading,'' in \emph{Proceedings of IEEE INFOCOM}, 2012.

\bibitem{SERENDIPITY_}
C.~Shi, V.~Lakafosis, M.~H. Ammar, and E.~W. Zegura, ``Serendipity: Enabling
  remote computing among intermittently connected mobile devices,'' in
  \emph{Proceedings of ACM MobiHoc}, 2012, pp. 145--154.

\bibitem{want2006introduction}
R.~Want, ``An introduction to rfid technology,'' \emph{Pervasive Computing,
  IEEE}, vol.~5, no.~1, pp. 25--33, 2006.

\bibitem{coskun2013professional}
V.~Coskun, K.~Ok, and B.~Ozdenizci, \emph{Professional NFC application
  development for android}.\hskip 1em plus 0.5em minus 0.4em\relax John Wiley
  \& Sons, 2013.

\bibitem{cite5}
B.~Hopkins, ``{Faster Data Transfer With Bluetooth and Contactless
  Communication},''
  \url{http://www.oracle.com/technetwork/articles/javame/nfc-bluetooth-142337.html},
  2009.

\bibitem{Chang:2010:NNP:1805161.1805165}
Y.-S. Chang, C.-L. Chang, Y.-S. Hung, and C.-T. Tsai, ``Ncash: Nfc
  phone-enabled personalized context awareness smart-home environment,''
  \emph{Cybern. Syst.}, vol.~41, no.~2, pp. 123--145, Feb. 2010.

\bibitem{cite7}
P.~Smith, ``{Comparing Low-Power Wireless Technologies},''
  \url{http://www.digikey.com/en/articles/techzone/2011/aug/comparing-low-power-wireless-technologies},
  2011.

\bibitem{ozdenizci2010nfc}
B.~Ozdenizci, M.~Aydin, V.~Coskun, and K.~Ok, ``Nfc research framework: a
  literature review and future research directions,'' in \emph{The 14th
  International Business Information Management Association (IBIMA) Conference.
  Istanbul, Turkey}, 2010.

\bibitem{cite9}
S.~CRAWFORD, ``{How Microsoft Surface Tabletop Works},''
  \url{http://computer.howstuffworks.com/microsoft-surface2.htm}, 2011.

\bibitem{7509388}
D.~Chatzopoulos, M.~Ahmadi, S.~Kosta, and P.~Hui, ``Openrp: a reputation
  middleware for opportunistic crowd computing,'' \emph{IEEE Communications
  Magazine}, vol.~54, no.~7, pp. 115--121, July 2016.

\bibitem{Fernando2013}
N.~Fernando, S.~W. Loke, and W.~Rahayu, ``Honeybee: A programming framework for
  mobile crowd computing,'' in \emph{Mobile and Ubiquitous Systems: Computing,
  Networking, and Services}.\hskip 1em plus 0.5em minus 0.4em\relax Springer
  Berlin Heidelberg, 2012, pp. 224--236.

\bibitem{Zhang:2014:CED:2685048.2685057}
I.~Zhang, A.~Szekeres, D.~Van~Aken, I.~Ackerman, S.~D. Gribble,
  A.~Krishnamurthy, and H.~M. Levy, ``Customizable and extensible deployment
  for mobile/cloud applications,'' in \emph{Proceedings of the 11th USENIX
  Conference on Operating Systems Design and Implementation}, ser. OSDI'14,
  2014, pp. 97--112.

\bibitem{Gordon:2015:AMA:2742647.2742649}
M.~S. Gordon, D.~K. Hong, P.~M. Chen, J.~Flinn, S.~Mahlke, and Z.~M. Mao,
  ``Accelerating mobile applications through flip-flop replication,'' in
  \emph{Proceedings of the 13th Annual International Conference on Mobile
  Systems, Applications, and Services}, ser. MobiSys '15, 2015, pp. 137--150.

\bibitem{6834974}
S.~Bouzefrane, A.~F.~B. Mostefa, F.~Houacine, and H.~Cagnon, ``Cloudlets
  authentication in nfc-based mobile computing,'' in \emph{Mobile Cloud
  Computing, Services, and Engineering (MobileCloud), 2014 2nd IEEE
  International Conference on}, April 2014, pp. 267--272.

\bibitem{7130873}
A.~Reiter and T.~Zefferer, ``Paving the way for security in cloud-based mobile
  augmentation systems,'' in \emph{Mobile Cloud Computing, Services, and
  Engineering (MobileCloud), 2015 3rd IEEE International Conference on}, March
  2015, pp. 89--98.

\bibitem{NQ}
J.~Somers, ``{The N Queens Problem: A Study in Optimization},'' \url{}, 2015.

\bibitem{RSA}
M.~Rouse, ``{RSA algorithm (Rivest-Shamir-Adleman)},''
  \url{http://searchsecurity.techtarget.com/definition/RSA}, Nov. 2014.

\end{thebibliography}

% that's all folks
\end{document}